\pdfoutput=1

\documentclass[english,aps,prd,superscriptaddress,preprintnumbers,floatfix,nofootinbib,12pt]{revtex4-1}
\usepackage{amsmath,amssymb,amsfonts}
\usepackage{amsmath}
\usepackage{graphicx}
\usepackage{orcidlink}

\usepackage{hyperref}
\hypersetup{colorlinks,linkcolor={blue},citecolor={blue},urlcolor={blue}} 
\usepackage{braket}
\usepackage[english]{babel}
\usepackage{bm}
\usepackage[utf8]{inputenc}

\usepackage{listings}

\usepackage[T1]{fontenc}
\usepackage{lmodern}
\usepackage{caption}
\usepackage{floatrow}
\usepackage{subcaption}
\usepackage{mathrsfs}
\usepackage{enumerate}
\usepackage{amsmath}
\usepackage{siunitx}

\newcommand{\GWth}{\text{GW}_{\text{th}}}
\usepackage[normalem]{ulem}
\usepackage{color}
\definecolor{nicered}{rgb}{0.8,0.15,0.15}

\def\beq{\begin{equation}}
\def\eeq{\end{equation}}
\def\be{\begin{equation}}
\def\ee{\end{equation}}
\def\bea{\begin{eqnarray}}
\def\eea{\end{eqnarray}}

\begin{document}

\title{Ultralight dark matter search in a large liquid scintillator detector}   
%\date{}
\author{Luis~A.~Delgadillo\orcidlink{0000-0003-2548-2299}} \email{ldelgadillof@ihep.ac.cn}
\affiliation{Institute of High Energy Physics, Chinese Academy of Sciences, Beijing 100049, China}  
\affiliation{Kaiping Neutrino Research Center, Guangdong 529386, China}  
\author{O.~G.~Miranda\orcidlink{0000-0002-0310-7060}} \email{omar.miranda@cinvestav.mx}
\affiliation{Departamento de F\'{\i}sica, Centro de Investigaci{\'o}n y de Estudios Avanzados del IPN Apdo. Postal 14-740 07000 Ciudad de M\'exico, Mexico}
\author{Hiroshi~Nunokawa\orcidlink{0000-0002-3369-0840}} \email{nunokawa@puc-rio.br}
\affiliation{Departamento de Física, Pontifícia Universidade Católica do Rio de Janeiro,\\
C.P. 38097, 22451-900, Rio de Janeiro, Brazil}

\begin{abstract}
\noindent
In this work, we investigate the phenomenological implications of ultralight scalar dark matter (ULDM) coupled to neutrinos. We focus on a large homogeneous liquid scintillator detector, analyzing the regime where ULDM oscillations lead to time-averaged distortions in neutrino oscillation probabilities. We derive sensitivity limits on the modulation parameters $\eta_{\Delta_{21}}$ and $\eta_{\Delta_{31}}$, which quantify ULDM-induced smearing effect in oscillations driven by solar ($\Delta m^2_{21}$) and atmospheric ($\Delta m^2_{31}$) mass-squared differences. We further demonstrate that ULDM interactions could produce a mild impact on both the determinations of the neutrino oscillation parameters and the neutrino mass ordering sensitivity. These results showcase the benefits of a large liquid scintillator detector as a powerful probe of neutrino-ULDM interactions via neutrino oscillations.

\end{abstract}

\maketitle

%\thispagestyle{empty}
%\vfill

%\newpage
%\setcounter{page}{1}

%%%%%%%%%%%%%%%%
%\newpage
%%%%%%%%%%%%%%%%

%%%%%%%%%%%%%%%%%%%%%%%%%%%%%%%%%%%%%%%%%%%%%%%%%%%%%%%%%%%%%%%%%%%%%%%%%%%%%%%%%%%%%%%%%%%%%%%%%%%%%%%%%%%%%%%%%%%%%%%

\section{Introduction}
\label{sec:intro}

Dark matter constitutes nearly one-fourth of the matter-energy content of the Universe, and the elucidation of its nature is both a fundamental challenge and one of the greatest open mysteries in physics. Among the  proposals to explain this phenomenon, one compelling candidate is an ultralight relativistic scalar field $\phi$, modeled through an appropriate scalar potential $V(\phi)$~\cite{Sahni:1999qe,Hu:2000ke,Matos:1999et,Hui:2016ltb, Khlopov:1985fch}. This theoretical framework is one of numerous other candidates spanning a wide mass spectrum. For instance, at the opposite mass extreme, the Weakly Interacting Massive Particle (WIMP) paradigm remains one of the best theoretically motivated candidates~\cite{Arbey:2021gdg}, despite extensive experimental searches having yielded no detectable signature so far. 

In this study, we focus specifically on the ultralight scalar field dark matter~(ULDM) candidate, since the dynamics of the ultralight scalar $\phi$ could give rise to significant and intriguing implications in particle physics. For instance, in the neutrino sector, owing to the small masses of neutrinos, their potential coupling with the ultralight scalar field background might substantially modify neutrino mass characteristics and flavor transformation probabilities. Given the unresolved nature of the neutrino mass generation mechanism, potential interactions between neutrinos and ULDM could shed light on phenomena beyond the Standard Model~(SM).

On the whole, the neutrino mass matrix is modified by its interactions with ULDM. The resulting neutrino oscillation probabilities become averaged when the ULDM oscillation period is much shorter than the detection period of time of an experiment. As a result, this averaging effect could be seen as a modification in the neutrino oscillation parameters when the experimental data is examined in the context of the conventional three-neutrino oscillation paradigm. For instance, studies of neutrino interactions between ultralight scalar dark matter and active neutrinos include~\cite{Berlin:2016woy, Barranco:2010xt, Reynoso:2016hjr, Brdar:2017kbt, Krnjaic:2017zlz, Capozzi:2018bps, Liao:2018byh, Farzan:2018pnk, Dev:2020kgz, Huang:2021kam, Losada:2021bxx, Huang:2022wmz, Cordero:2022fwb, Sen:2023uga, Lin:2023xyk, Graf:2023dzf, Martinez-Mirave:2024dmw, Lambiase:2025twn, Ge:2024ftz, Cheek:2025kks, Chao:2025sao}.

In this work, we consider the possibility of neutrino interactions with ultralight dark matter at a liquid scintillator detector setup similar to JUNO. For its basic characteristics, namely: fiducial volume, baseline, reactor power, and so on (for brevity, hereafter, we refer to such a liquid scintillator detector configuration simply as JUNO in this work). We focus on the scenario of time-averaged oscillations mediated by the ULDM field, which could lead to an additional smearing effect, adding to that of the experimental energy resolution. We assess the resulting sensitivities, the impact on the determination of the standard oscillation parameters, and the potential implications of this scenario for determining the neutrino mass ordering. Previous work on this subject has been performed in Refs.~\cite{Krnjaic:2017zlz, Losada:2021bxx}. However, here we make a more detailed analysis that takes into account the most recent information about JUNO, such as energy resolution and smearing function and study the impact of the ultra-light scalar field-neutrino interaction on the standard oscillation parameters and mass ordering.

The manuscript is organized as follows. In Section~\ref{sec:framework}, we introduce the phenomenology and theoretical framework of neutrino oscillations in the presence of ultralight dark matter. Section~\ref{sec:setup} describes the experimental configuration, assumptions, and simulation of the JUNO-like setup considered in this analysis. Subsequently, Section~\ref{sec:analysis} details the methodology for quantifying the statistical significance of the neutrino-ultralight-scalar-field dark matter scenario. In Section~\ref{sec:results}, we present potential signatures of ultralight dark matter through neutrino oscillations at a large liquid scintillator detector; specifically, we examine the impact on neutrino oscillation parameter determination and the neutrino mass ordering sensitivity. Finally, Section~\ref{sec:conclusions} provides our conclusions and discusses future assessments in this direction.

%%%%%%%%%%%%%%%%%%%%%%%%%%%%%%%%%%%%%%%%%%%%%%%%%%%%%%%%%%%%%%%%%%%%%%%%%%%%%%%%%%%%%%%%%%%%%%%%%%%%%%%%%%%%%%%%%%%%%%%

\section{Neutrino oscillations in the presence of ultralight scalar field}
\label{sec:framework}

Considering the ultralight scalar field $\phi$ as a dark matter (DM) candidate, specifically ultralight dark matter (ULDM), with potential $V(\phi) \simeq m_\phi^2 \phi^2$, the corresponding field is given by~\cite{Suarez:2013iw, Ferreira:2020fam}
\begin{equation}
   \phi(t,\mathbf{X}) \simeq \frac{\sqrt{2 \rho_{\phi, \odot}}}{m_\phi} \sin\big(m_\phi (t - \langle v_\phi \rangle \mathbf{X}) \big) \simeq \phi_0 \sin\big(m_\phi (t - \langle v_\phi \rangle \mathbf{X}) \big) \;,
\end{equation}
where $\rho_{\phi, \odot}$ denotes the local scalar field density, $m_\phi$ the corresponding scalar field mass, $\langle v_\phi \rangle \sim 10^{-3}$ is the virialized velocity in the Milky Way (MW), and $\mathbf{X}=(x, y, z)$ represents the spatial coordinates. The ultralight scalar field can be treated as a classical coherent wave provided its occupation number per de Broglie volume is large, $\langle \mathcal{N}_\phi \rangle \sim \rho_{\text{DM},\odot}/(m_\phi^4 \langle v_\phi \rangle^3)\gg 1$, which is satisfied in the Milky Way halo for the local dark matter density $\rho_{\text{DM}, \odot}\approx 0.3\ \text{GeV}/\text{cm}^3$ and virial velocity $\langle v_\phi \rangle\sim 10^{-3}$~\cite{Arvanitaki:2014faa}. For instance, for a scalar field mass $m_\phi \sim 10^{-15}$ eV, the occupation number is $\langle \mathcal{N}_\phi \rangle \sim 10^{63}$. Besides, in the non-relativistic regime, once the Hubble friction becomes negligible ($H \sim m_\phi$), the classical field equation~\cite{Cordero:2022fwb} \begin{equation}
\ddot{\phi}+3H\dot{\phi}+\frac{dV(\phi)}{d\phi}= 0\;,
\end{equation} results in an oscillatory solution. In a matter-dominated Universe, the energy density goes as $\rho(T)\propto T^4$. Hence, the Hubble parameter scales as
\begin{equation}
    H^2 = \frac{8 \pi G}{3} \rho(T) \sim \frac{T^4}{M_{\text{Pl}}^2}\;,
\end{equation}
where $M_{\text{Pl}}$ is the Planck mass and $T$ is the corresponding temperature. Setting $H \sim m_\phi$ gives the oscillation temperature,
\begin{equation}
T_{\text{osc}}\sim \sqrt{M_{\text{Pl}} m_{\phi}} \sim \text{MeV} \left(\frac{M_{\text{Pl}}}{10^{28}~\text{eV}}\right)^{1/2} \left(\frac{m_{\phi}}{10^{-15}~\text{eV}}\right)^{1/2}\;.
\end{equation} Thus, for the ultralight scalar field masses considered in this study ($m_\phi \gtrsim 10^{-21}$ eV), this transition occurs at early epochs. Hence, locally, the ultralight scalar field is approximated as $ \phi(t,\mathbf{X}) \simeq \phi(t) = \phi_0 \sin(m_\phi t)$, being $\phi_0 \equiv \sqrt{2 \rho_{\phi,\odot}}/m_\phi$ the scalar field amplitude. Such ultralight scalars (with masses down to $m_{\phi} \sim 10^{-33}$ eV) appear in certain theoretical frameworks~\cite{Arvanitaki:2009fg,Cicoli:2021gss}.~\footnote{Searches for ultralight scalars as dark matter candidates include atomic clocks~\cite{Arvanitaki:2014faa}, pulsar timing array~\cite{Smarra:2024kvv}, resonant-mass detectors~\cite{Arvanitaki:2015iga}, atomic gravitational wave detectors~\cite{Arvanitaki:2016fyj}, as well as optical cavities~\cite{Campbell:2021mnu}.} 
For instance, the Lagrangian for a scalar field $\phi$ coupled to active neutrino species $\nu_\alpha$ is given by the term~\cite{Krnjaic:2017zlz}
\begin{equation}
    \label{eq:lag}
    -\mathcal{L}_{\nu\phi} \supset(m_{\alpha\beta} + \hat{y}_{\alpha\beta} \phi) \bar{\nu}_\alpha \nu_\beta \,,
\end{equation}
where $m_{\alpha\beta}$ is the neutrino mass matrix and $\hat{y}_{\alpha\beta}$ is the Yukawa coupling matrix governing the interaction between the active neutrinos and the scalar field. In this scenario, the neutrino--ultralight-scalar dark matter coupling therefore contributes to the neutrino mass.~\footnote{For other scenarios where ULDM accounts for the dominant contribution to neutrino masses, see Refs.~\cite{Sen:2023uga, Ge:2024ftz, Cheek:2025kks}.} Consequently, a perturbation from the scalar field background $\phi$ modifies the neutrino mass matrix to
\begin{equation}
    \hat{m}_{\alpha \beta} = m_{\alpha \beta} +\hat{y}_{\alpha \beta} \phi\;,
\end{equation}
and the effective mass-squared difference as~\cite{Losada:2021bxx}
\begin{equation}
    \Delta\hat m^2_{jk} = \Delta m^2_{jk}+2(m_j\hat y_{jj}-m_k\hat y_{kk})\phi+{\cal O}\left(\hat y^2\phi^2\right) \equiv\Delta m^2_{jk}\left[1+2\eta_{\Delta_{jk}}\sin(m_\phi t)\right]\;.
\end{equation}
Moreover, the corresponding neutrino mass values are modulated by the ultralight scalar field amplitude as~\cite{Cordero:2022fwb}~\footnote{For discussions on the impact of active neutrinos on ultralight scalar field dynamics and other cosmological implications, we refer the reader to Refs.~\cite{Cordero:2022fwb, Sen:2023uga}.}
\begin{equation}
\label{eta}
   \eta_{\Delta_{jk}} = \phi_0 \frac{ m_j \hat{y}_{jj} - m_k \hat{y}_{kk}}{ \Delta m_{jk}^2}= \frac{ \sqrt{2 \rho_{\phi,\odot}} (m_j \hat{y}_{jj} - m_k \hat{y}_{kk}) }{m_{\phi} \Delta m_{jk}^2}\;.
\end{equation}
Hence, the neutrino conversion probability acquires a modification due to shifts in the diagonal neutrino mass terms. 

In this work, we focus on the diagonal $\hat{y}_{jj}$ couplings that are present in the term $\eta_{\Delta_{jk}}\propto m_j \hat{y}_{jj} - m_k \hat{y}_{kk}$, as they directly contribute to the effective Hamiltonian in Eq.~(5). A complete analysis including off-diagonal terms elements would require a re-diagonalization of the full neutrino mass matrix, which we leave for future work.
In particular, neutrino oscillation experiments that determine vacuum oscillations can extract an averaged value~\cite{Losada:2021bxx}
\begin{equation}
\label{eq:fastdjk}
\begin{split}
    &\big \langle\sin^2[\Delta\hat m^2_{jk}L/(4E)]\big \rangle \equiv 
\frac{1}{\tau_\phi}\int_0^{\tau_\phi} dt\ \sin^2\Big( x_{Ejk}[1+2\eta_{\Delta_{jk}}\sin(m_\phi t)]\Big) =\\ 
 & ~ \sin^{2}\left(x_{Ejk}\right) + 2x_{Ejk}^{2}\eta_{\Delta_{jk}}^{2}\cos\left(2x_{Ejk}\right) + \mathcal{O}\left(x_{Ejk}^{4}\eta_{\Delta_{jk}}^{4}\right),~x_{Ejk} \equiv \frac{\Delta m^2_{jk} L}{4E} \;,
\end{split}
\end{equation}
where $\tau_\phi$ indicates the period of the time variation of the scalar field, $L$ is the baseline (or the distance traveled by neutrino) and $E$ is the neutrino energy.
In this work we will consider the case where the oscillation frequency 
$\omega_{\text{osc}} = m_\phi$ is much faster than the time scales relevant for neutrino's trajectory~\cite{Cordero:2022fwb}, leading to the approximation $\langle \cos(2m_\phi L) \rangle_{T} \approx 0$.
Therefore, the effective $\eta_{\Delta_{jk}}$ parameters induce an energy smearing effect that adds to the intrinsic energy resolution of the experiment. Thus, a JUNO-like experiment provides a unique opportunity to explore the sensitivity to both the $\eta_{\Delta_{31}}$ and $\eta_{\Delta_{21}}$ parameters.

One of the characteristic features of the scenario considered in this work is that, as shown in Fig.~1 of \cite{Krnjaic:2017zlz}, the neutrino oscillation signal exhibits damping of the oscillatory behaviors~\cite{JUNO:2021ydg}, an effect similar to decoherence~\cite{Lisi:2000zt,Gago:2000qc,Coloma:2018idr} or neutrino decay~\cite{Gonzalez-Garcia:2008mgl,Abrahao:2015rba}.

%%%%%%%%%%%%%%%%%%%%%%%%%%%%%%%%%%%%%%%%%%%%%%%%%%%%%%%%%%%%%%%%%%%%%%%%%%%%%%%%%%%%%%%%%%%%%%%%%%%%%%%%%%%%%%%%%%%%%%%

\section{Experimental configuration}
\label{sec:setup}

An example of such a large liquid scintillator detector is, for instance, the Jiangmen Underground Neutrino Observatory (JUNO)~\cite{JUNO:2015zny,JUNO:2021vlw}, located in China. The primary scientific goal of the JUNO experiment is to determine the neutrino mass ordering~\cite{Zhan:2009rs, Li:2013zyd, Bilenky:2017rzu, Forero:2021lax, Parke:2024xre, JUNO:2024jaw}, as well as precision  measurements of some of the oscillation parameters~\cite{JUNO:2022mxj}. Recently, the JUNO collaboration reported its first results on the detector's performance~\cite{JUNO:2025fpc} and a precision measurement of the solar mixing parameters~\cite{JUNO:2025gmd}. However, other physics potentials at a JUNO-like experimental configuration include studies related to solar neutrinos~\cite{JUNO:2023zty}, geoneutrinos~\cite{Han:2015roa, JUNO:2025sfc}, atmospheric neutrinos~\cite{JUNO:2021tll, Liu:2025fry, Birkenfeld:2025cxe}, supernova neutrinos~\cite{JUNO:2022lpc}, proton decay~\cite{JUNO:2022qgr}, neutron decay~\cite{JUNO:2024pur}, MeV dark matter searches~\cite{JUNO:2023vyz}, long-baseline neutrino synergies~\cite{Cabrera:2020ksc, Goswami:2025wla}, neutrino non-standard interactions (NSI)~\cite{Li:2014mlo, Liao:2017awz, Martinez-Mirave:2021cvh}, neutrino Lorentz and CPT violation~\cite{Li:2014rya, Barenboim:2023krl}, ultralight scalar dark matter~\cite{Krnjaic:2017zlz, Losada:2021bxx, Losada:2022uvr, Losada:2023zap}, as well as other Beyond Standard Model (BSM) scenarios~\cite{Wang:2020uvi, Huber:2021xpx, Basto-Gonzalez:2021aus, deGouvea:2021uvg, Lucente:2022esm}.

In this work, we employ the GLoBES software~\cite{Huber:2004ka,Huber:2007ji} to simulate a large liquid scintillator detector similar to JUNO. We consider a \SI{20}{kton} liquid scintillator detector located at \SI{52.5}{km} from the reactor core. We have assumed an exposure of $6.5~\text{years} \times 26.6~\GWth$ of reactor thermal power. We also assume $90\%$ signal  detection efficiency and employ the Huber-Mueller flux model~\cite{Huber:2011wv, Mueller:2011nm} to calculate the predicted signal event rates. A constant matter density of $\rho = \SI{2.45}{g/cm^3}$ is considered following \cite{JUNO:2022mxj}. Regarding expected background contaminants, we adopt the total expected background rates from Ref.~\cite{JUNO:2024jaw}. In this analysis, we consider 410 bins with a constant bin size of 20 keV for the neutrino energy spectrum reconstruction, spanning from 1.8 MeV to 10 MeV. Furthermore, in this work, the energy resolution function, which relates the true neutrino energy ($E$) to the reconstructed energy ($E^{\prime}$), follows a Gaussian distribution~\cite{Huber:2002mx}:
\begin{equation}
    R(E,E^{\prime}) = \frac{1}{\sqrt{2\pi}  \sigma_{R}(E)} 
    \exp \left( -\frac{(E - E')^2}{2 \sigma_{R}^2(E)} \right),
\end{equation}
where $\sigma_{R}(E) = \beta \sqrt{E / \si{MeV}} \:\si{MeV}$ and $\beta = 0.03$~\cite{JUNO:2015sjr}.
%~\cite{Li:2013zyd}. 
This specification corresponds to a $3\%$ energy resolution at $E = \SI{1}{MeV}$.

\begin{figure}[H]
\includegraphics[scale=0.635]{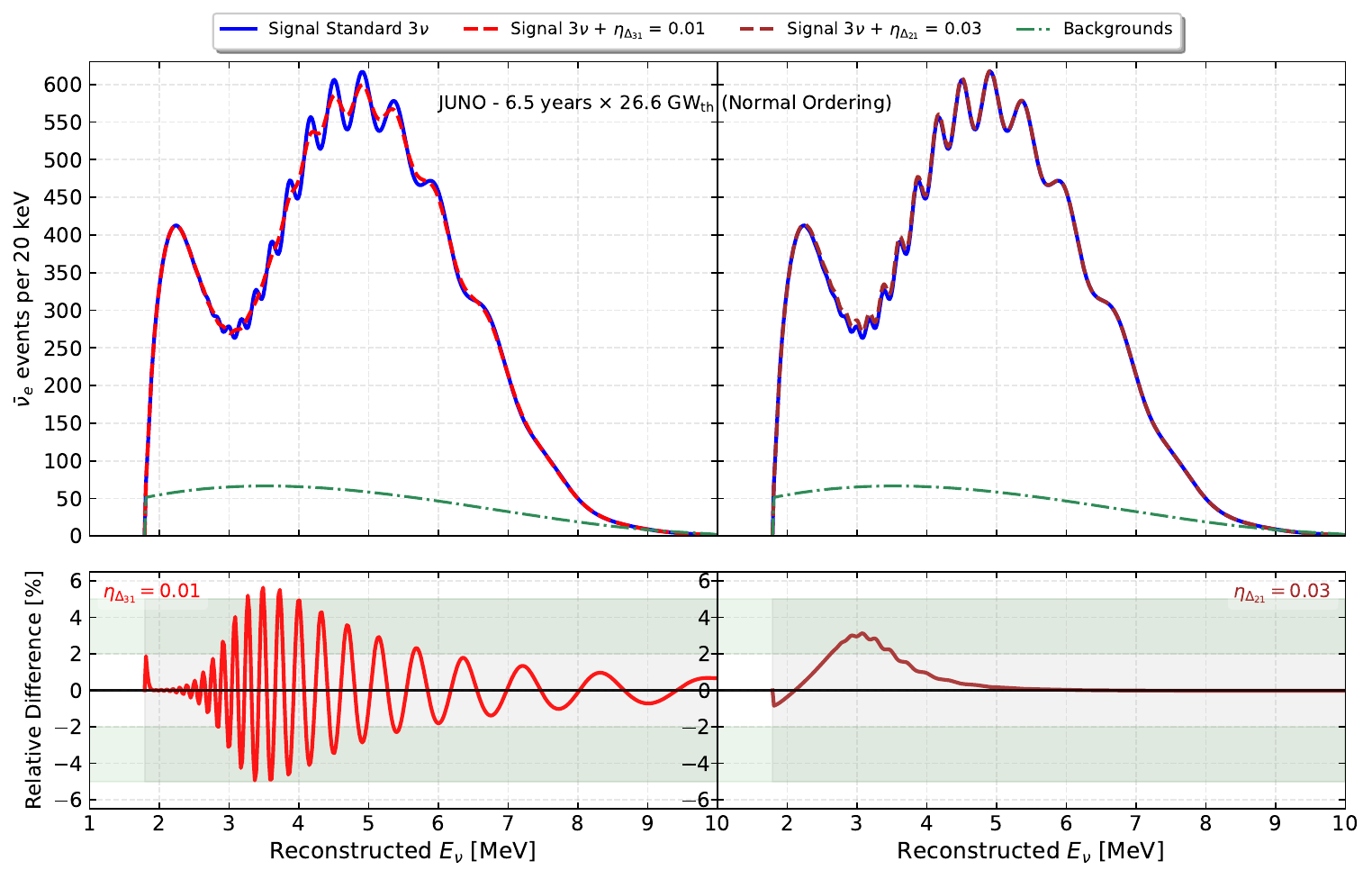}
		 \caption{
         Upper panels: Reconstructed spectra at the JUNO-like setup showing the impact of non-zero $\eta_{\Delta_{31}}$ (left panel)
         and $\eta_{\Delta_{21}}$ (right panel). The blue solid line represents the spectrum within the standard three-neutrino ($3\nu$) oscillation framework. The green dashed line shows the expected background. The red and (brown) dashed lines represent the spectra including ultralight scalar field effects assuming $\eta_{\Delta_{31}} = 0.01$ and ($\eta_{\Delta_{21}} = 0.03$), respectively. 
         Lower panels: Corresponding relative differences of the event rates in the presence of the scalar interaction to the standard oscillation rate. See the text for a detailed explanation.}
          
  \label{fig:f1}
\end{figure}
In Fig.~\ref{fig:f1}, we show in the upper panels the reconstructed neutrino energy spectra at the JUNO-like configuration with and without the ULDM effect obtained by using the mixing parameters for the normal ordering (NO) of neutrino masses found in Table \ref{tab:osc} in the next section. The blue-solid line represents the predicted spectrum under the standard three-neutrino oscillation paradigm. Moreover, the green-dashed line indicates the total expected background contribution at the large liquid scintillator detector. The upper left panel displays the modified spectrum from ultralight scalar modulations on the neutrino mass-squared difference $\Delta m^2_{31}$ with parameter $\eta_{\Delta_{31}} = 0.01$ (red-dashed line). The upper right panel shows the modified spectrum from ultralight scalar modulations on $\Delta m^2_{21}$ with parameter $\eta_{\Delta_{21}} = 0.03$ (brown-dashed line). 
In order to visualize more clearly the impact of the ULDM effect on the energy spectra, in the lower panels we show the corresponding relative differences of the event rates in the presence of the scalar interaction to the standard oscillation rate. These comparative spectra highlight the impact of ultralight scalar interactions on the observed energy distribution relative to the standard $3\nu$ oscillation case.

%%%%%%%%%%%%%%%%%%%%%%%%%%%%%%%%%%%%%%%%%%%%%%%%%%%%%%%%%%%%%%%%%%%%%%%%%%%%%%%%%%%%%%%%%%%%%%%%%%%%%%%%%%%%%%%%%%%%%%%

\section{Analysis procedure}
\label{sec:analysis}
We set a least-squares analysis to quantify the statistical significance of the neutrino ultralight scalar field dark matter scenario. The total $\chi^2$ function is given as~\cite{Huber:2002mx}
\begin{equation}
    \chi^2 = \tilde{\chi}^2 + \chi^2_{\text{prior}},
\end{equation}
where $\tilde{\chi}^2$ is based on the number of events computed for the $\bar{\nu}_e \rightarrow \bar{\nu}_e$ channel and is defined as in Ref.~\cite{Huber:2002mx}:
\begin{equation}
\begin{split}
    &\tilde{\chi}^2= \min_{\xi_{j}} \Bigg[  \sum _{i}^{n_{\text{bin}}} 2 \Bigg\{ N_{i,\text{test}}^{3 \nu+{\eta}}( \Xi, \Pi, \{\xi_{j}\})-N_{i,\text{true}}^{3\nu} + N_{i,\text{true}}^{3\nu} \log \frac{N_{i,\text{true}}^{3\nu}}{N_{i,\text{test}}^{3 \nu+\eta}( \Xi, \Pi, \{\xi_{j}\})} \Bigg\} \\
    &~~~~~~~~~~~~~~~ + \sum_{j}^{n_{\text{syst}}} \Big(\frac{\xi_{j}}{\sigma_{j}}\Big)^2 \Bigg]\;.
\end{split}
\end{equation}
Here, $N_{i, \text{true}}^{3\nu}$ represents the simulated events in the $i$-th energy bin under the standard three-neutrino oscillation framework, while $N_{i, \text{test}}^{3\nu + \eta}( \Xi, \Pi, \{\xi_{j}\})$ denotes the predicted events in the $i$-th bin, including the $\eta_{\Delta_{jk}}$ parameters (analyzed one at a time). The set $\Xi = \{\theta_{12}, \theta_{13}, \Delta m_{21}^2, \Delta m^2_{31}\}$ contains the neutrino oscillation parameters, and $\Pi = \{\eta_{\Delta_{21}}, \eta_{\Delta_{31}} \}$ comprises parameters for neutrino interactions with the ultralight scalar field. Systematic uncertainties are incorporated via nuisance parameters $\{\xi_{j}\}$. We assume a 5\% signal normalization uncertainty, a 20\% background normalization uncertainty, and a 3\% energy calibration uncertainty for both signal and background, consistent with Ref.~\cite{Porto-Silva:2020gma}. The simulated events are generated using the assumed \emph{true} neutrino oscillation parameters taken from Salas \emph{et al.}~\cite{deSalas:2020pgw}, listed in Table~\ref{tab:osc}.

\begin{table}[ht]
\centering
\caption{\label{tab:osc}Standard oscillation parameters used in our analysis~\cite{deSalas:2020pgw}. Unless otherwise specified, we consider the NO throughout this study.}
\begin{tabular}{c  c}
\hline \hline
Oscillation parameter & Best fit NO  \\
\hline 
$\theta_{12}$ & 34.3$^{\circ}$ \\
$\theta_{13}$ &  8.53$^{\circ}$ \\
$\Delta m^2_{21}$ [10$^{-5}$~eV$^2$] & 7.5  \\ 
$|\Delta m_{31}^2|$ [10$^{-3}$~eV$^2$] & 2.55  \\ 
\hline \hline
\end{tabular}
\end{table} 
External constraints on the standard oscillation parameters are implemented through Gaussian priors~\cite{Huber:2002mx}:
\begin{equation}
    \chi^2_{\text{prior}} = \sum_{k}^{n_{\text{priors}}}  \frac{\big( \Xi_{k,\text{true}} - \Xi_{k,\text{test}} \big)^2}{\sigma^2_{k}}\,.
\end{equation}
The central values $\Xi_{k,\text{true}}$ are fixed to their best-fit values from Ref.~\cite{deSalas:2020pgw} (assuming normal ordering), while the uncertainties $\sigma_k$ correspond to the $1\sigma$ allowed ranges. These priors are marginalized over the standard oscillation parameters $k$ of interest.

%%%%%%%%%%%%%%%%%%%%%%%%%%%%%%%%%%%%%%%%%%%%%%%%%%%%%%%%%%%%%%%%%%%%%%%%%%%%%%%%%%%%%%%%%%%%%%%%%%%%%%%%%%%%%%%%%%%%%%%

\section{Results}
\label{sec:results}
In this section, we consider modulations of the ultralight scalar field in the regime where \( \tau_{\nu} \ll \tau_{\phi} \ll \tau_{\text{exp}} \) (time-averaged modulations). Here, \( \tau_{\nu} = L / c \) is the neutrino time of flight, \( \tau_{\phi} = 0.41 \times (10^{-14}  \,\text{eV} / m_{\phi}) \, \text{seconds} \) is the characteristic modulation period of the ultralight scalar field~\cite{Dev:2020kgz}, and \( \tau_{\text{exp}} \) denotes the exposure time of the experiment. Under these circumstances, oscillation effects from mixing angles or mass splittings are too rapid to be observed, but an averaging effect on oscillation probabilities could be detected~\cite{Krnjaic:2017zlz,Dev:2020kgz,Losada:2021bxx}.

Regarding the ultralight scalar field mass sensitivity at the large liquid scintillator (LLS) detector similar to JUNO, the baseline is \( L = 52.5~\text{km} \), yielding a neutrino time of flight \( \tau_{\nu} = 1.8 \times 10^{-4}~\text{seconds} \). For an exposure time of \( \tau_{\text{exp}} = 6.5~\text{years} \) and a scalar field modulation period \( \tau_{\phi} \approx 1~\text{year} \), we expect the ultralight scalar field mass sensitivity to be 
\begin{equation}
 3.0 \times 10^{-23}\  \text{eV} \lesssim m_{\phi}^{\rm{LLS}}\lesssim 2.0 \times 10^{-11}\  \text{eV} \;.   
\end{equation}

\subsection{Sensitivity to the $\eta_{\Delta_{jk}}$ parameters via time-averaged modulations}
\begin{figure}[H]
		\begin{subfigure}[h]{0.45\textwidth}
			\caption{  }
			\label{fg1}
\includegraphics[width=\textwidth]{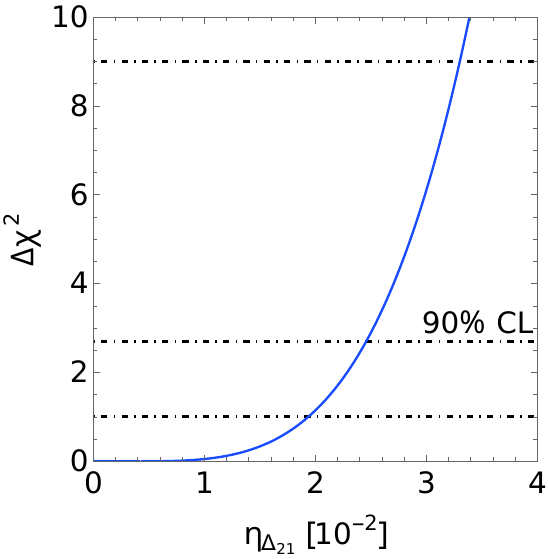}
		\end{subfigure}
		\hfill
		\begin{subfigure}[h]{0.45\textwidth}
			\caption{}
			\label{fg2}
			\includegraphics[width=\textwidth]{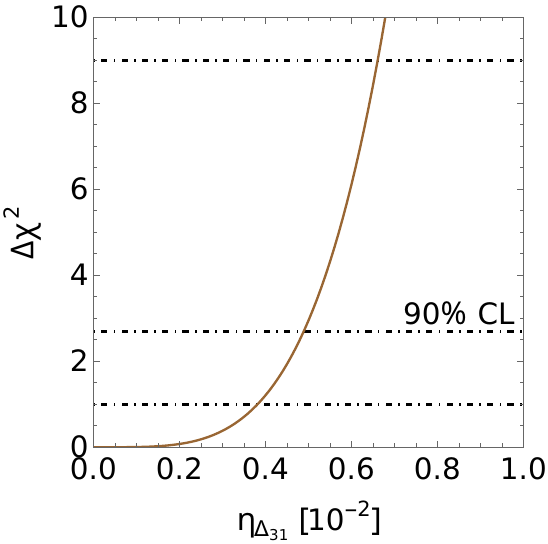}
		\end{subfigure}
		\hfill	
 \caption{Projected $\Delta \chi^2$ sensitivity to the ultralight scalar field parameters $\eta_{\Delta_{21}}$ (left panel) and $\eta_{\Delta_{31}}$ (right panel) from mass-squared modulations at the JUNO-like configuration. See the text for a detailed explanation.}
  \label{fig:f2}
\end{figure}
In Fig.~\ref{fig:f2}, we present the expected sensitivities to the ultralight scalar field parameters $\eta_{\Delta_{jk}}$ via time-averaged modulations at JUNO. The left panel shows the expected sensitivity to $\eta_{\Delta_{21}}$, while the right panel displays the projected sensitivity to $\eta_{\Delta_{31}}$, both ultralight scalar field parameters. Therefore, the corresponding 90\% CL limits ($\Delta \chi^2 = 2.7$) on the ultralight scalar field parameters at this detector configuration are $\eta_{\Delta_{21}} \lesssim 2.5 \times 10^{-2}$ and $\eta_{\Delta_{31}} \lesssim 0.5 \times 10^{-2}$.

A preliminary assessment of the projected sensitivities to the ultralight scalar field parameters $\eta_{\Delta_{jk}}$ via time-averaged modulations at JUNO was performed in Ref.~\cite{Krnjaic:2017zlz}. Additionally, sensitivities to these parameters via time-resolved modulations at JUNO were evaluated in Ref.~\cite{Losada:2021bxx}. For illustration purposes, in order to relate the sensitivity of the phenomenological parameters $\eta_{\Delta_{jk}}$ to the Yukawa couplings $\hat{y}_{jk}$, let us consider a scalar field density $\rho_{\phi} = 0.1 \rho_{\rm{DM}} \simeq 10^{-12}$ eV$^4$ (with local DM density $\rho_{\rm{DM}, \odot} \simeq 10^{5} \rho_{\rm{DM}}$~\cite{Planck:2018vyg}) and mass $m_\phi \simeq 10^{-21}$ eV.~\footnote{For scalar field masses lighter than $m_\phi \lesssim 10^{-21}$ eV, the scalar field density $\rho_{\phi}$ cannot constitute the total cosmological DM density~\cite{Cordero:2022fwb}.} Therefore, from the limits on the $\eta_{\Delta_{jk}}$ parameters, we can constrain the corresponding Yukawa couplings $\hat{y}_{jk} \propto m_j \hat{y}_{jj} - m_k \hat{y}_{kk} \propto \eta_{\Delta_{jk}}$ (see Eq.~\ref{eta}). 

Hence, the projected 90\% CL bounds on neutrino–ultralight-scalar-field couplings for a large liquid scintillator detector similar to JUNO are as follows:
\begin{equation}
    \begin{split}
&\hat{y}_{21} \lesssim 3 \times 10^{-22}~\big(\eta_{\Delta_{21}}/2.5\times 10^{-2} \big)\big(\Delta m^2_{21} / 7.5\times 10 ^{-5} {\rm{eV^2}} \big)\big( m_\phi/10^{-21} {\rm{eV}} \big)\big( \rho_{\phi}/ 0.1 \rho_{\rm{DM}, \odot}[ {\rm{eV}^4} ] \big)\;, \\
&\hat{y}_{31} \lesssim 4\times 10^{-22}~\big(\eta_{\Delta_{31}}/0.5\times 10^{-2} \big)\big(\Delta m^2_{31} / 2.5\times 10 ^{-3} {\rm{eV^2}} \big)\big( m_\phi/10^{-21} {\rm{eV}} \big)\big( \rho_{\phi}/ 0.1 \rho_{\rm{DM}, \odot}[ {\rm{eV}^4} ] \big)\;. \\
\end{split}
\end{equation}

\begin{figure}[H]
\includegraphics[scale=0.7]{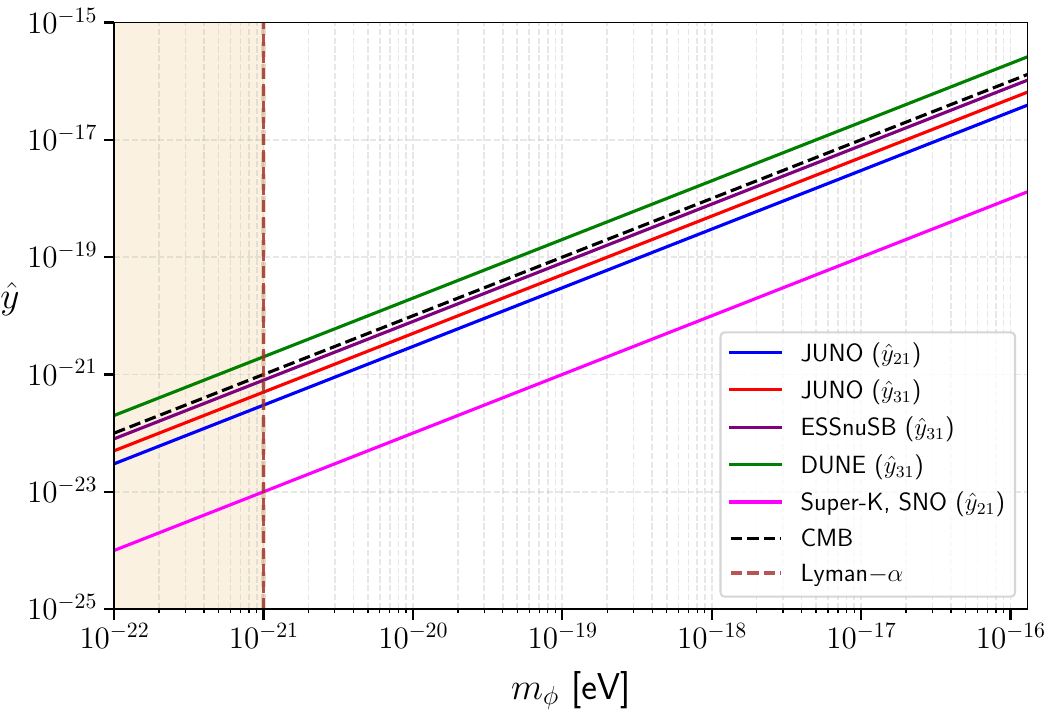}
		 \caption{Limits and sensitivity regions in the $\hat{y}-m_\phi$ plane. The black dashed line represents the excluded limits from the CMB measurements on the total sum of the neutrino masses~\cite{Planck:2018vyg}. In the brown-shaded region, ultralight scalar dark matter lighter than $m_\phi \sim 10^{-21}\ \mathrm{eV}$ is in tension with Lyman-$\alpha$ forest observations~\cite{Berlin:2016woy}. Furthermore, solid lines indicate the expected sensitivities for the neutrino--ultralight dark matter scenario via time-averaged modulations from the JUNO-like setup (this work), DUNE~\cite{Dev:2020kgz}, and ESSnuSB~\cite{Cordero:2022fwb}, along with current constraints from solar neutrino measurements (Super-K and SNO)~\cite{Berlin:2016woy}. Here, these limits were derived under the assumption that ultralight dark matter constitutes 10\% of the total dark matter abundance.} 
  \label{fig:f3}
\end{figure}

In Fig.~\ref{fig:f3}, we show the constraints and projected sensitivities for the neutrino--ultralight dark matter scenario in the $\hat{y}-m_\phi$ parameter space. The black dashed line indicates the exclusion limits from CMB measurements of the total neutrino mass sum~\cite{Planck:2018vyg}, $\hat{y}_{\rm{CMB}} \lesssim 10^{-22}~(m_\phi/10^{-22} \ \mathrm{eV})$. Solid lines denote projected sensitivities due to time-average modulations from JUNO (this work), DUNE~\cite{Dev:2020kgz}, and ESSnuSB~\cite{Cordero:2022fwb}, alongside existing constraints from Super-Kamiokande (Super-K) and SNO~\cite{Berlin:2016woy}. When compared with current constraints, we can notice that $\hat{y}_{21}$ sensitivity is not competitive with current bounds~\cite{Berlin:2016woy} while the projected $\hat{y}_{31}$ sensitivity is more promising.
However, ultralight scalar dark matter below $m_\phi \sim 10^{-21}\ \mathrm{eV}$ is disfavored by Lyman-$\alpha$ forest data~\cite{Berlin:2016woy}, as shown in the brown-shaded region. Here, all derived limits consider the case when ultralight dark matter comprises 10\% of the total dark matter density. In addition, ultra-faint dwarf (UFD) galaxies constraint $m_\phi  \gtrsim  10^{-19}\ \mathrm{eV}$~\cite{Dalal:2022rmp}, while if dark matter is produced after inflation, $m_\phi \gtrsim 10^{-19}\ \mathrm{eV}$ accounting for up to 20\% of the total dark matter density~\cite{Amin:2022nlh}.

\subsection{Correlations among $\eta_{\Delta_{jk}}$ and the mass squared differences $\Delta m_{jk}^2$}
\begin{figure}[H]
		\begin{subfigure}[h]{0.495\textwidth}
			\caption{  }
			\label{fh1}
\includegraphics[width=\textwidth]{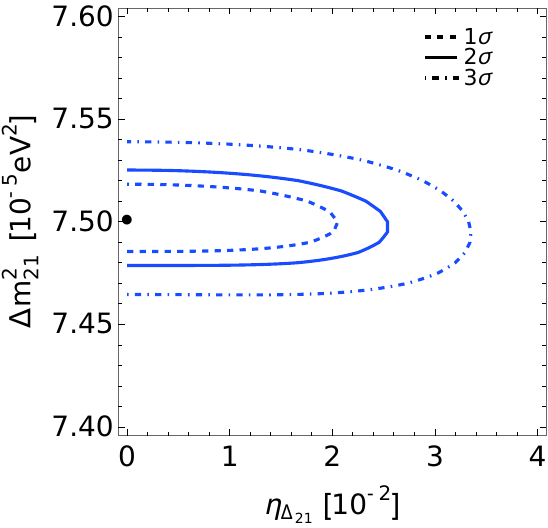}
		\end{subfigure}
		\hfill
		\begin{subfigure}[h]{0.495\textwidth}
			\caption{}
			\label{fh2}
			\includegraphics[width=\textwidth]{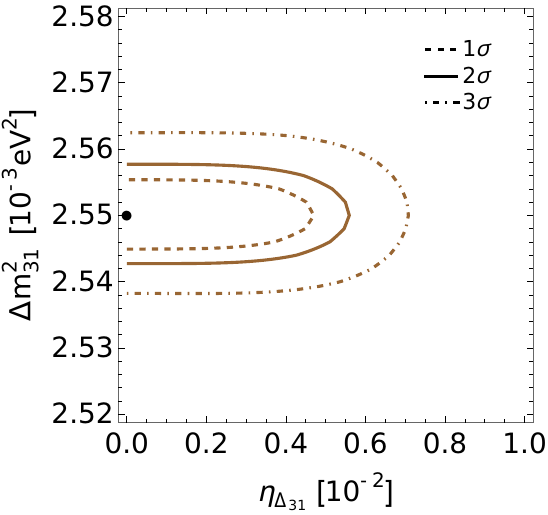}
		\end{subfigure}
		\hfill	
 \caption{Correlations among the ultralight scalar field parameters $\eta_{\Delta_{21}}$ ($\eta_{\Delta_{31}}$) from the corresponding solar (atmospheric) mass-squared modulations for a JUNO-like experiment. Contours enclosed by the dashed, solid, and dot-dashed lines denote the $1\sigma$, $2\sigma$, and $3\sigma$ sensitivity regions, accordingly. See the text for a detailed explanation.}
  \label{fig:f4}
\end{figure}

In Fig.~\ref{fig:f4}, we show the correlations among the ultralight scalar field parameters $\eta_{\Delta_{jk}}$ with the corresponding mass-squared differences $\Delta m^2 _{jk}$ via time-averaged modulations at the JUNO-like setup. The left panel displays the correlation between the $\eta_{\Delta_{21}}$ and the solar $\Delta m^2_{21}$ mass-squared splitting, while right panel shows the corresponding correlation among $\eta_{\Delta_{31}}$ and the atmospheric $\Delta m^2_{31}$ mass-squared difference. The contours enclosed by the dashed, solid, and dot-dashed lines denote the $1\sigma$, $2\sigma$, and $3\sigma$ sensitivity regions, respectively. Regarding the $\eta_{\Delta_{21}}$ parameter, discrimination with respect to the standard three-flavour oscillation scenario at $3\sigma$ is possible if $\eta_{\Delta_{21}} > 3.4 \times 10^{-2}$ (left panel of Fig.~\ref{fig:f4}). While for the $\eta_{\Delta_{31}}$ parameter, distinction with respect to the standard three-flavour oscillation paradigm at $3\sigma$ is achieved if $\eta_{\Delta_{31}} > 7.5 \times 10^{-3}$ (right panel of Fig.~\ref{fig:f4}).

\subsection{Impact of $\eta_{\Delta_{jk}}$ on the determination of the oscillation parameters}
\begin{figure}[H]
		\begin{subfigure}[h]{0.495\textwidth}
			\caption{  }
			\label{fj1}
\includegraphics[width=\textwidth]{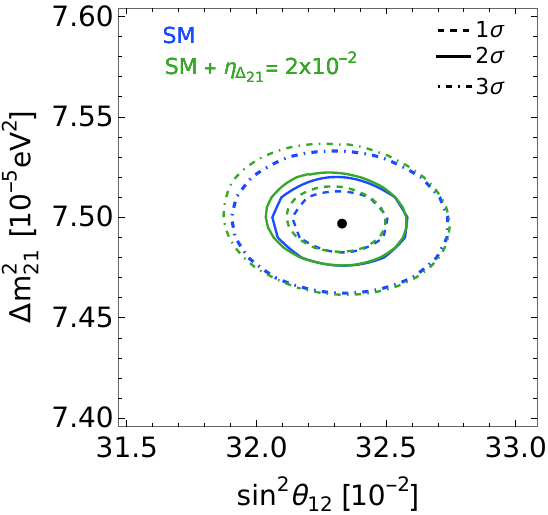}
		\end{subfigure}
		\hfill
		\begin{subfigure}[h]{0.495\textwidth}
			\caption{}
			\label{fj2}
			\includegraphics[width=\textwidth]{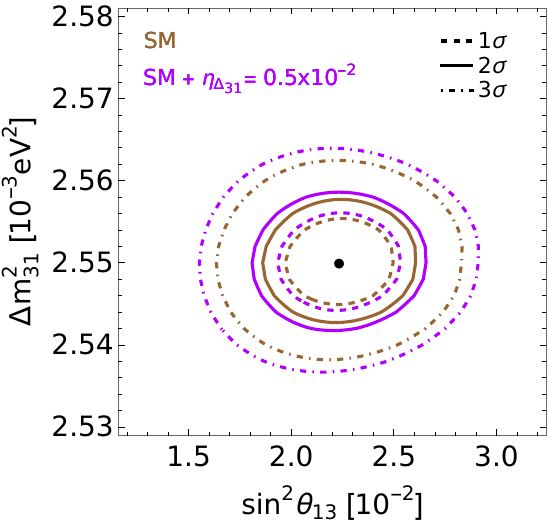}
		\end{subfigure}
		\hfill	
 \caption{Impact of ultralight scalar field parameters $\eta_{\Delta_{jk}}$ on the determination of oscillation parameters: $\sin^2\theta_{12}$ vs.~$\Delta m^2_{21}$ (left panel) and $\sin^2\theta_{13}$ vs.~$\Delta m^2_{31}$ (right panel). The standard three-neutrino scenario (SM) serves as reference. Green and purple lines show the effect when $\eta_{\Delta_{jk}}$ parameters are included in both simulation input and fit (SM + $\eta_{\Delta_{jk}}$). Input values were fixed to $\eta_{\Delta_{21}} = 2 \times 10^{-2}$, and $\eta_{\Delta_{31}} = 5 \times 10^{-3}$, accordingly. See the text for a detailed explanation.}
  \label{fig:f5}
\end{figure}
In Fig.~\ref{fig:f5}, we present the impact of the ultralight scalar field parameters $\eta_{\Delta_{jk}}$ on the determination of the oscillation parameters: $\sin^2\theta_{12}$ vs.~$\Delta m^2_{21}$ (left panel), and $\sin^2\theta_{13}$ vs.~$\Delta m^2_{31}$ (right panel). The standard $3\nu$ oscillation paradigm (SM) is shown as reference. For the solar parameters ($\sin^2\theta_{12}$, $\Delta m^2_{21}$), we marginalized over $\theta_{13}$ and $\Delta m^2_{31}$ within their 1$\sigma$ uncertainties from Ref.~\cite{deSalas:2020pgw}. Conversely, for the atmospheric parameters ($\sin^2\theta_{13}$, $\Delta m^2_{31}$), we marginalized over $\theta_{12}$ and $\Delta m^2_{21}$ within their 1$\sigma$ uncertainties. The green (purple) lines present the impact of including the $\eta_{\Delta_{jk}}$ parameters in both the input simulation and the fit, SM + $\eta_{\Delta_{jk}}$. We set either $\eta_{\Delta_{21}} = 2 \times 10^{-2}$ or ($\eta_{\Delta_{31}} = 5 \times 10^{-3}$) as input values corresponding to its projected $1\sigma$ sensitivity (Fig.~\ref{fig:f2}), and allow them to vary in the fit.~\footnote{We have also considered the case where only the $\eta_{\Delta_{jk}}$ parameters were allowed to vary in the fit while maintaining the standard $3\nu$ framework as input, but their impact on the determination of oscillation parameters was marginal.} While the impact of the $\eta_{\Delta_{21}}$ parameter on the determination of the solar oscillation parameters ($\sin^2\theta_{12}$, $\Delta m^2_{21}$) is negligible, the $\eta_{\Delta_{31}}$ parameter moderately impacts the atmospheric parameters ($\sin^2\theta_{13}$, $\Delta m^2_{31}$), reducing the precision of their measurements at the aforementioned detector configuration.

\subsection{Impact of $\eta_{\Delta_{jk}}$ on the mass ordering determination}

In this subsection, we investigate the effects of the ultralight scalar field on neutrino mass ordering determination at the large liquid scintillator configuration. Specifically, we quantify how interactions between neutrinos and the ultralight scalar field, parameterized through $\eta_{\Delta_{jk}}$, influence the sensitivity to distinguish between normal and wrong mass orderings. Our analysis focuses on the degradation of $\Delta\chi^2$ sensitivity when these new physics parameters are included in both simulation and fitting procedures.
\begin{figure}[H]
\includegraphics[scale=0.9]{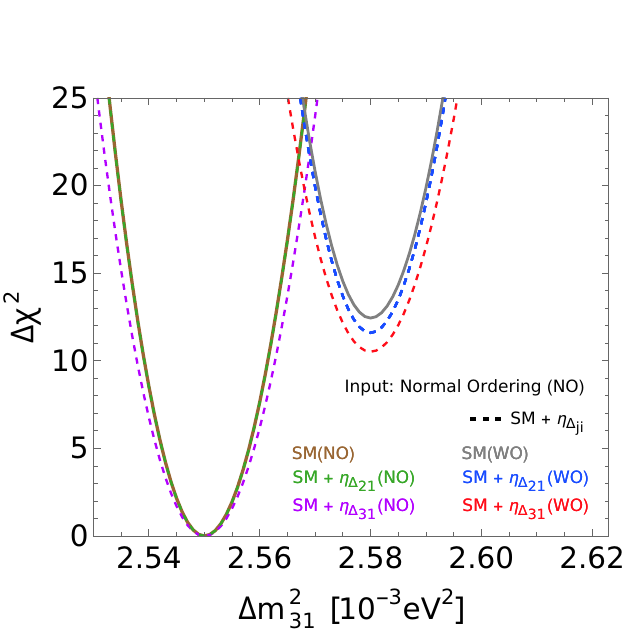}
		 \caption{The corresponding $\Delta \chi^2$ profile as a function of the fitted $\Delta m_{31}^2$ value. The solid brown and gray curves correspond to cases where the fit assumes the normal (right) and inverted (wrong) mass ordering, respectively, under standard oscillations. The input scenario assumes normal mass ordering (NO). Dashed lines green, purple, (blue) and (red) represent cases where $\eta_{\Delta_{jk}}$ parameters are included in both input simulation and fit, for both correct NO and wrong (WO) mass orderings. See text for detailed explanation.} 
  \label{fig:f6}
\end{figure}
In Fig.~\ref{fig:f6}, we consider normal mass ordering NO, as the input scenario, corresponding to $\Delta m_{31}^2 = 2.55 \times 10^{-3}~\mathrm{eV}^2$ (brown solid line). Additionally, we include a wrong ordering (WO) case with $\Delta m_{31}^2 = 2.58 \times 10^{-3}~\mathrm{eV}^2$ (gray solid line) for comparison. The $\Delta \chi^2$ between the correct (NO) and wrong (WO) mass orderings is:
\begin{equation}
    \Delta \chi^2_{\mathrm{NO-WO}} \simeq 13.
\end{equation}
The dashed lines show the impact of including the $\eta_{\Delta_{jk}}$ parameters in both the input simulation and the fit. We set either $\eta_{\Delta_{21}} = 2 \times 10^{-2}$ or $\eta_{\Delta_{31}} = 5 \times 10^{-3}$ as input values corresponding to its expected sensitivity at JUNO (Fig.~\ref{fig:f2}), and allow them to vary in the fit.~\footnote{The impact of including the $\eta_{\Delta_{jk}}$ parameters only in the fit (not in the input) was found to be marginal, with $\Delta\chi^2 < 1$.}

While the $\eta_{\Delta_{31}}$ parameter moderately affects the sensitivity to neutrino mass ordering determination, $\eta_{\Delta_{21}}$ exhibits a milder impact. For instance, when including $\eta_{\Delta_{21}}$, the sensitivity decreases by approximately one unit, yielding $\Delta \chi^2_{\mathrm{NO-WO}} -\eta_{\Delta_{21}} \simeq 12$. In contrast, including $\eta_{\Delta_{31}}$ reduces the sensitivity by about 2.5 units, resulting in $\Delta \chi^2_{\mathrm{NO-WO}} -\eta_{\Delta_{31}} \simeq 10.5$. This reduction limits the sensitivity reach for normal mass ordering determination.

\begin{figure}[H]
\includegraphics[scale=0.5]{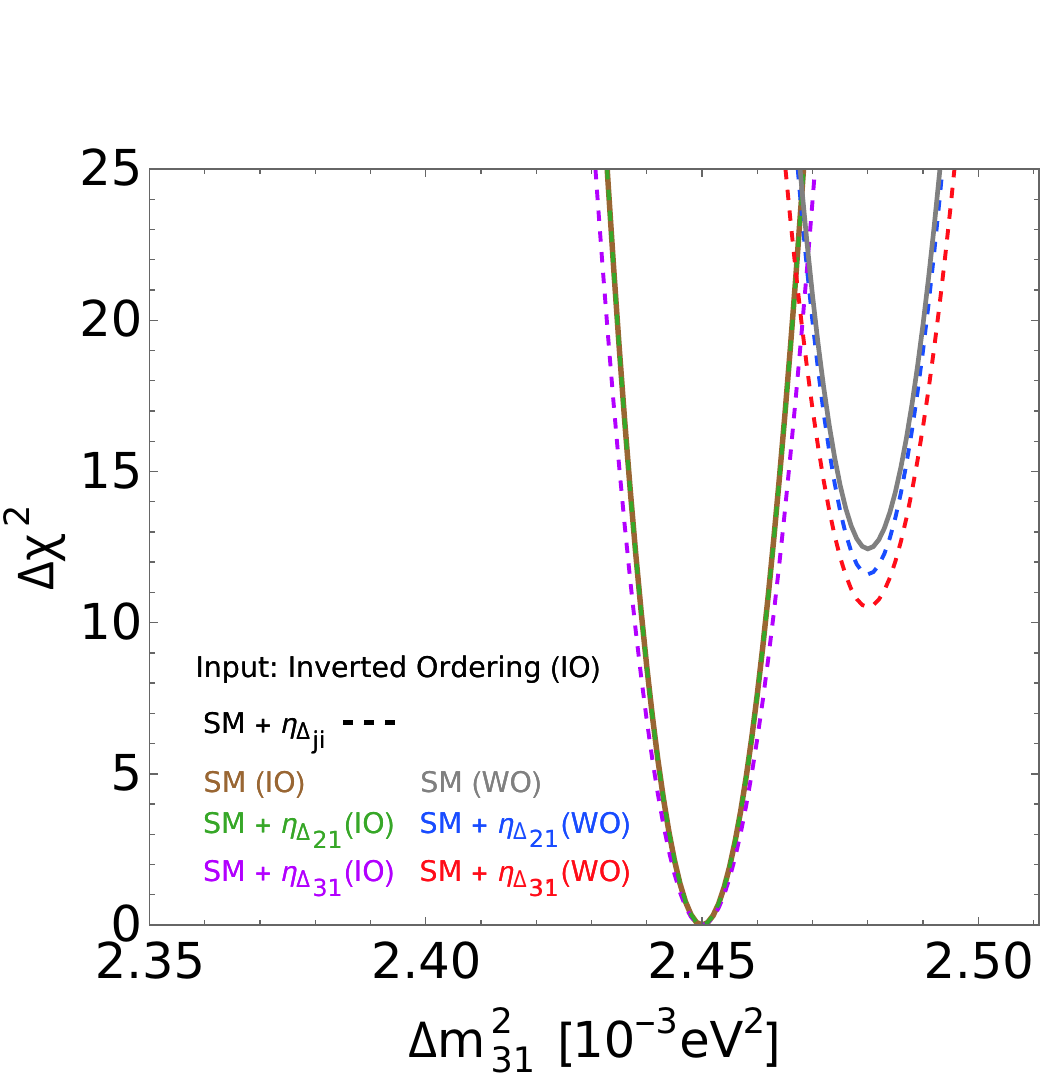}
		 \caption{The corresponding $\Delta \chi^2$ profile as a function of the fitted $\Delta m_{31}^2$ value. The solid brown and gray curves correspond to cases where the fit assumes the inverted (right) and normal (wrong) mass ordering, respectively, under standard oscillations. The input scenario assumes inverted mass ordering (IO). Dashed lines green, purple, (blue) and (red) represent cases where $\eta_{\Delta_{jk}}$ parameters are included in both input simulation and fit, for both correct IO and wrong (WO) mass orderings. See text for detailed explanation.} 
  \label{fig:f7}
\end{figure}
In Fig.~\ref{fig:f7}, we consider normal mass ordering IO, as the input scenario, corresponding to $\Delta m_{31}^2 = 2.45 \times 10^{-3}~\mathrm{eV}^2$ (brown solid line). Additionally, we include a wrong ordering (WO) case with $\Delta m_{31}^2 = 2.48 \times 10^{-3}~\mathrm{eV}^2$ (gray solid line) for comparison. Besides, dashed lines show the impact of including the $\eta_{\Delta_{jk}}$ parameters in both the input simulation and the fit. We set either $\eta_{\Delta_{21}} = 2 \times 10^{-2}$ or $\eta_{\Delta_{31}} = 5 \times 10^{-3}$ as input values corresponding to its expected sensitivity at JUNO (Fig.~\ref{fig:f2}), and allow them to vary in the fit. The corresponding $\Delta \chi^2$ between the correct (IO) and wrong (WO) mass orderings is:
\begin{equation}
    \Delta \chi^2_{\mathrm{IO-WO}} \simeq 12.
\end{equation}
However, a precise quantitative assessment will require a detailed implementation of liquid scintillator non-linear effects (LSNL). Nevertheless, this provides a qualitative indication of how such ultralight scalar interactions with neutrinos may affect neutrino mass ordering determination at such type of configuration.

%%%%%%%%%%%%%%%%%%%%%%%%%%%%%%%%%%%%%%%%%%%%%%%%%%%%%%%%%%%%%%%%%%%%%%%%%%%%%%%%%%%%%%%%%%%%%%%%%%%%%%%%%%%%%%%%%%%%%%%

\section{Conclusions}
\label{sec:conclusions}
In this paper, we explored the impact of ultralight scalar dark matter (ULDM) on neutrino oscillations at a JUNO-like liquid scintillator detector. We show that couplings between neutrinos and a coherently oscillating ULDM field $\phi$ introduce time-averaged modulations in neutrino mass-squared differences, parameterized by $\eta_{\Delta_{jk}}$. For ULDM masses $m_\phi \sim 10^{-23}\text{--}10^{-11}$ eV, such configuration could set 90\% CL limits of $\eta_{\Delta_{21}} < 2.5 \times 10^{-2}$ and $\eta_{\Delta_{31}} < 5 \times 10^{-3}$, accordingly (Fig.~\ref{fig:f2}). These translate to bounds on the neutrino-ULDM Yukawa couplings $\hat{y}_{21} \lesssim 3 \times 10^{-22}$ and $\hat{y}_{31} \lesssim 4 \times 10^{-22}$ for $m_\phi = 10^{-21}$ eV and $\rho_\phi = 0.1 \rho_{{\rm DM},\odot}$, surpassing current constraints from the CMB and complementing projections from DUNE and ESSnuSB experiments (Fig.~\ref{fig:f3}).

Crucially, ULDM-neutrino interactions introduce spectral distortions that mimic energy smearing. Nevertheless, the impact of such interactions posses mild correlations with the corresponding mass-squared differences: $\Delta m^2_{21}$ and $\Delta m^2_{31}$ (Fig.~\ref{fig:f4}). When unaccounted for these effects, the determination of the solar mixing angle, $\sin^2\theta_{12}$, remains robust, while for the reactor mixing angle, $\sin^2\theta_{13}$, a moderate reduction of its precision could be obtained (Fig.~\ref{fig:f5}). However, the sensitivity to the neutrino mass ordering could be degraded by $\Delta\chi^2 \simeq 2.5$, when $\eta_{\Delta_{31}} \sim 5 \times 10^{-3}$ (Figs.~\ref{fig:f6} and \ref{fig:f7}).

Our findings underscore the importance of including ULDM interactions in precision neutrino studies. Future work in this direction should incorporate detailed liquid scintillator non-nonlinearities, precise knowledge of backgrounds as well as additional neutrino sources. The intersection of neutrino physics and dark matter remains a fertile frontier, and experiments like JUNO will play a leading role in constraining or discovering neutrino couplings to ultralight dark sectors.

\section*{Acknowledgments}
The work of L.~A.~D. has been supported by the Kaiping Neutrino Research Center (KNRC) and IHEP-CAS. 
H.~N. was supported by the Brazilian funding agencies CNPq and CAPES. The work of O. G. M. has been supported by SNII (Sistema Nacional de Investigadoras e Investigadores, Mexico). We thank the anonymous referee for the comments and suggestions that have helped us to improve our manuscript.

This paper represents the views of the authors and should not be considered a JUNO collaboration paper.

\bibliography{biblio.bib}
\end{document}